\documentclass[floatfix,aps,prl,twocolumn,amsmath,amssymb,groupedaddress]{revtex4}
\usepackage{graphicx}
\begin{document}


\title{High-contrast modulation of light with light by control of
surface plasmon polariton wave coupling}


\author{A. V. Krasavin}
\email[]{avk@soton.ac.uk}
\author{K. F. MacDonald}
\author{N. I. Zheludev}
\affiliation{School of Physics and Astronomy, University of
Southampton, SO17 1BJ, United Kingdom}
\author{A. V. Zayats}
\affiliation{School of Mathematics and Physics, Queen's University
of Belfast, BT7 1NN, United Kingdom}


\date{\today}

\begin{abstract}
We have demonstrated a new mechanism for modulating light with
light by controlling the efficiency with which light is coupled
into a plasmon polariton wave. An optical fluence of 15 mJ/cm$^2$
in the control channel is sufficient to achieve nearly a 10-fold
intensity modulation of the signal beam reflected from a
Glass/MgF$_2$/Ga structure. The mechanism depends on a nanoscale
light-induced structural transformation in the gallium layer and
has transient switching times of the order of a few tens of
nanoseconds. It offers high modulation contrast for signals in the
visible and near infrared spectral ranges.
\end{abstract}

\pacs{Here should be pacs}

\maketitle

Surface plasmon polariton (SPP) waves, i.e. surface
electromagnetic excitations coupled with electrons at a
metal-dielectric interface \cite{Raether}, are attracting
increasing attention as a potential new type of information
carrier for future highly integrated photonic devices. A range of
very promising nanostructures that direct and guide SPP waves and
allow for sub-wavelength structural elements on plasmonic "chips"
are now being investigated \cite{Barnes, az-jopa-03}. However, it
will not be possible to speak about "plasmonics" in the same way
that we speak about "photonics" until techniques for active
manipulation of SPP signals are developed. Recent theoretical
analysis shows that active switching of plasmonic signals should
be possible through a stimulation-induced nanoscale structural
transformation in the waveguide material \cite{Krasavin}. Here, in
a further exploration of this idea, we report for the first time
that a light-induced nanoscale structural transformation can be
used to control the efficiency with which electromagnetic
radiation is coupled into SPP waves, and thus modulate the
intensity of optical signals.

SPP waves propagate along the interface between a metal and a
dielectric. As the dispersion relation for SPP's is different from
that for light, it is only possible to couple light into an SPP
wave on a smooth surface by using a matching device such as a
grating, or a prism placed at the interface \cite{Raether}.
Effective coupling is possible only in an optimized regime, i.e.
for a particular grating period or for a particular angle of
incidence through the prism. The coupling efficiency also depends
on the dielectric characteristics of the materials in a very thin
region around the interface. That part of the light wave which is
not coupled into the SPP wave is reflected from the interface. Our
modulation concept is based on the idea that by changing the
dielectric characteristics of the metal at the interface through a
light-induced structural transformation, one can drive the system
away from the resonant coupling conditions and thus exercise
control over the intensity of the wave reflected from the
interface.
\begin{figure*}
    \includegraphics[width=160mm]{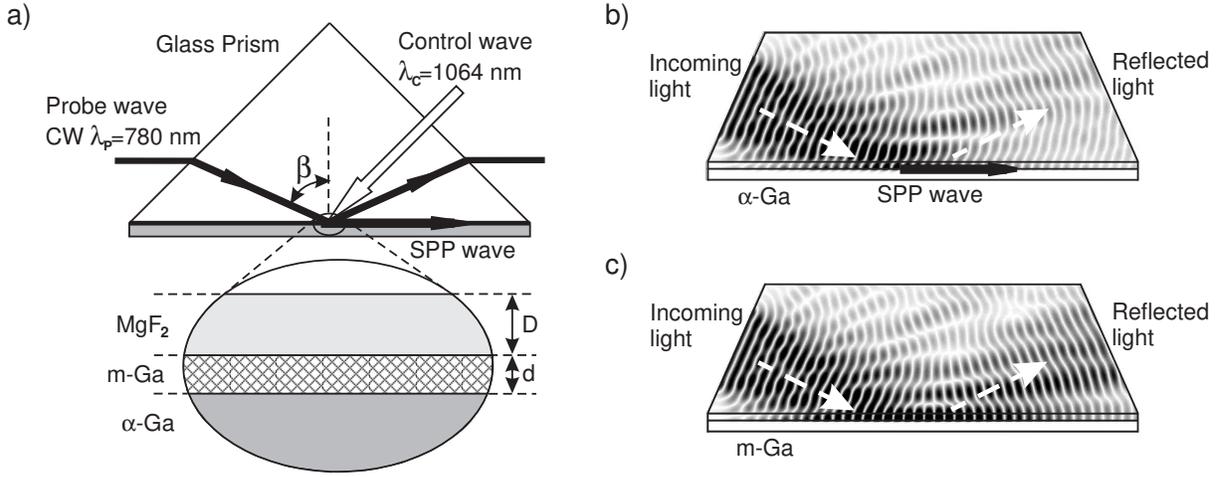}\\
  \caption{(a) Arrangements for modulation of light with light and control over light-SPP wave coupling in a Glass/MgF$_2$/Ga structure
  in the Otto configuration. (b,c) Field distributions in the vicinity of the Glass/MgF$_2$/Ga structure calculated for (b)
  $\alpha$- and (c) m- phases of gallium.}\label{Fig1}
\end{figure*}

Gallium is a uniquely suitable material to realize this concept.
It is known for its polymorphism \cite{Defrain} and
$\alpha$-gallium, the stable "ground-state" phase \cite{Gong}, has
a very low melting point, 29.8$^\circ$C, and is partially covalent
bound. The optical properties of $\alpha$-Ga and those of the more
metallic phases, which are metastable under normal conditions, are
very different. The properties of the metastable phases are
similar to those of the highly metallic liquid phase. In terms of
the dielectric coefficients at a wavelength of 780 nm,
$|\varepsilon_{liquid}|/|\varepsilon_\alpha|\sim5$. A metastable
metallic phase (quasi-melt) may be achieved at the interface by
simple heating, or by light absorption through a non-thermal
"optical melting" mechanism based on the destabilization of the
optically excited covalent bonding structure, which only affects a
few atomic layers of the material at an interface \cite{Zheludev}.
Such structural transformations have already been shown to provide
photonic functionality, offering, for example, all optical
switching at normal reflection from bulk interfaces \cite{Albanis}
and nanoparticle films \cite{MacDonald}. Here we demonstrate
experimentally that the use of a stimulated structural
transformation to control the coupling between light and SPP waves
provides modulation contrast more than one order of magnitude
higher than at bulk interfaces (at normal incidence) and about two
orders of magnitude higher than for nanoparticle films.
\begin{figure}[h]
    \includegraphics[width=75mm]{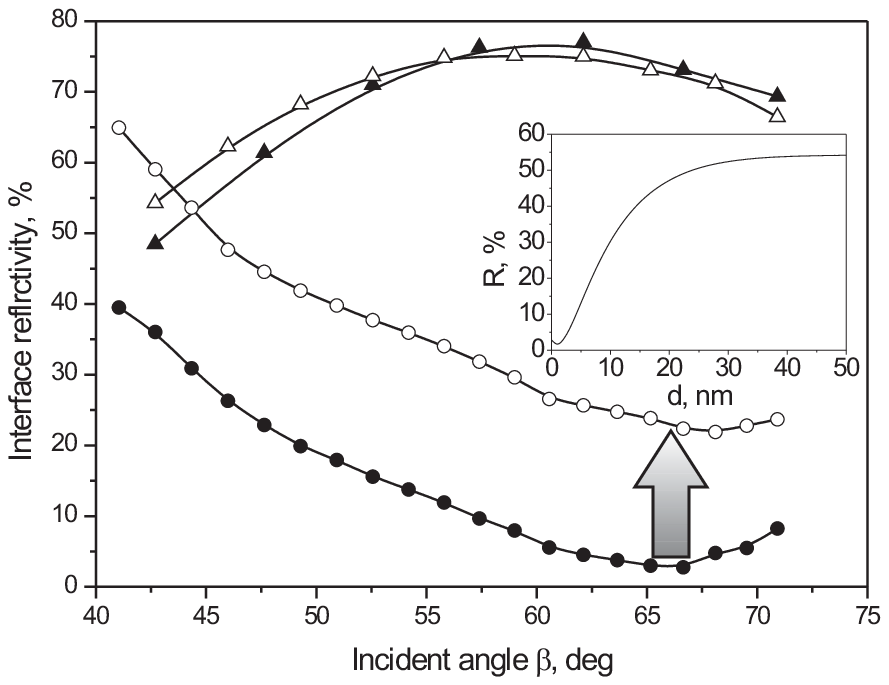}\\
  \caption{Angular dependence of reflectivity from the Glass/MgF$_2$/Ga structure.
  Filled and open symbols represent $\alpha$- and metallic gallium phases respectively. Triangles and circles denote s- and p-polarizations of the incident light.
The vertical arrow shows the reflectivity change that can be
achieved by a complete transformation of the gallium film from the
$\alpha$-phase into the metallic phase. The inset shows the
theoretical dependence of reflectivity on the thickness $d$ of the
metallic m-phase of gallium.}\label{Fig2}
\end{figure}

In our experiments we used the attenuated total reflection device
in Otto configuration \cite{Raether} for coupling light to an SPP
wave. Here gallium is interfaced with a BK7 glass prism covered
with a MgF$_2$ film of thickness $D$ = 185 nm (Fig. \ref{Fig1} a).
A gallium film was prepared on the prism by squeezing a bead of
the liquid metal then solidifying it. A light wave undergoing
total internal reflection on the Glass/MgF$_2$ interface is
efficiently coupled to an SPP wave at the MgF$_2$/Ga interface at
a particular (resonant) angle of incidence where the photon wave
vector in glass is equal to the SPP wave vector on the MgF$_2$/Ga
interface. A continuous wave diode laser operating at
$\lambda_p$=780nm was used as a probe source (Fig. \ref{Fig1} a).
The dependencies of the reflectivity of the Glass/MgF$_2$/Ga
structure on incident angle for $s$ and $p$ polarizations are
presented in Fig. \ref{Fig2} for two different Ga phases. The
reflectivity of the structure has a clear minimum for the $p$
polarization, but not for the $s$ polarization of the incident
light. The reflectivity minimum corresponds to the resonant
conditions for coupling light into the SPP wave, which can only be
achieved for the $p$ polarization \cite{Raether}.

The resonant SPP coupling is illustrated by numerical modelling of
the electromagnetic fields in the interface area, as presented in
Fig.1 (b) and (c). The gray scale in these images represents the
amplitude of the magnetic component of the field (darker for
higher amplitude). Image (b) shows the case for high coupling
efficiency and low reflectance, under resonant conditions of SPP
excitation ($\beta$=66$^\circ$), when the energy of the incoming
wave is converted into the SPP wave and rapidly damped because of
high SPP wave losses ($\alpha$-Ga phase at the interface). Image
(c) shows the case for low coupling efficiency and high
reflectance (gallium is in the metallic phase at the interface).

The inset in Fig. \ref{Fig2} shows the interface reflectivity at
$\beta$ = 66$^\circ$, the resonant angle for excitation of the SPP
wave on the MgF$_2$/Ga interface, as a function of increasing
thickness $d$ of the metallic layer, calculated using data on
dielectric coefficients from \cite{Kofman} and \cite{Teshev}. From
here one can see that by changing Ga from the $\alpha$-phase to
the m-phase in a layer only a few nanometers thick, the coupling
efficiency, and thus the reflected beam intensity at the resonant
incident angle, can be changed dramatically.
\begin{figure}
    \includegraphics[width=75mm]{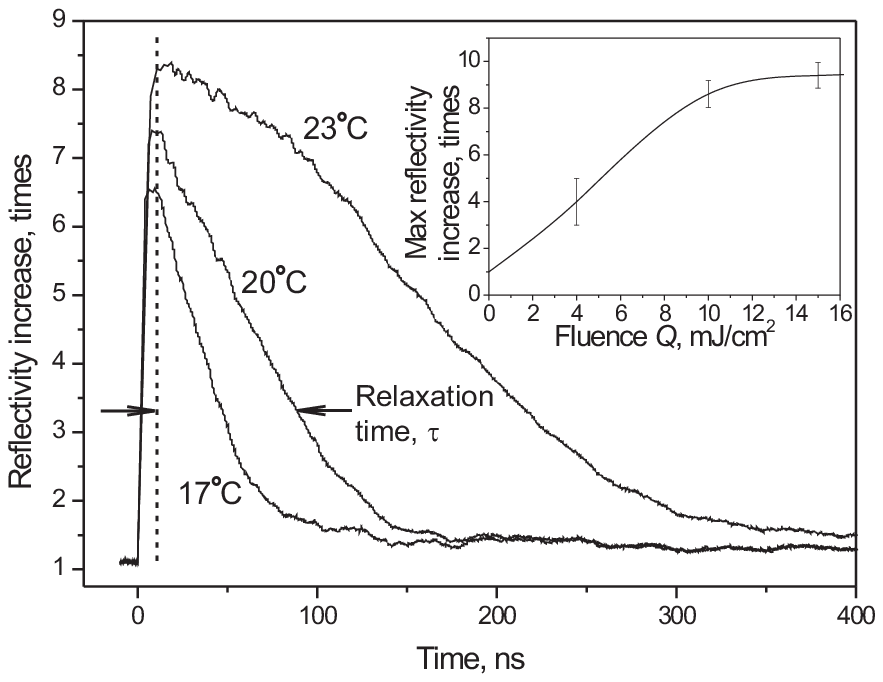}\\
  \caption{Transient reflectivity of the Glass/MgF$_2$/Ga structure following 6ns impulse excitation
  at a wavelength of 1.06 $\mu$m (Q=15 mW/cm$^2$) for various structure temperatures. The angle of incidence is 66$^\circ$. The inset shows the
  dependence of the maximum reflectivity increase on control wave fluence at $28^\circ$C.}\label{Fig3}
\end{figure}
\begin{figure}
    \includegraphics[width=75mm]{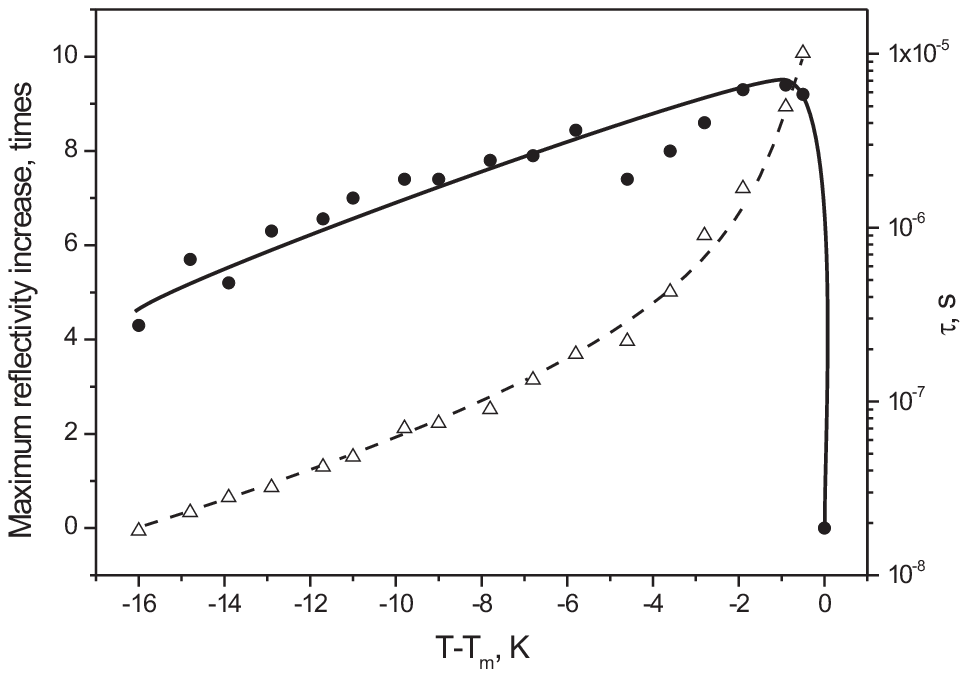}\\
  \caption{Temperature dependence of the maximum reflectivity increase ($\bullet$) and corresponding relaxation time ($\triangle$) at Q=15 mJ/cm$^2$.}\label{Fig4}
\end{figure}

To demonstrate light-by-light modulation of the reflected probe
beam via a structural transformation from the $\alpha$-phase to
the metallic phase in the gallium film, we introduced a channel
for optical excitation of the interface with a Nd:YAG laser,
generating 6 ns pulses at $\lambda_c$=1064 nm with the repetition
rate of 20 Hz (Fig. \ref{Fig1} a). The control and probe laser
spots were overlapped on the interface. Stimulation with the
control laser leads to an immediate increase in the reflected
probe intensity $R$. At an excitation fluence of $Q$=15 mJ/cm$^2$
the obtained reflectivity increase $R/R_{off}$ reaches 9.4, where
the $R_{off}$ is the reflectivity of the $\alpha$-Ga when the pump
laser is off. This significant change in the intensity of the
reflected wave corresponds only to about a 20$\%$ decrease in the
efficiency of coupling into the SPP wave. The magnitude of the
effect increases with the fluence up to about 15 mJ/cm$^2$ and
then saturates (insert in Fig \ref{Fig3}). This behavior may be
explained as follows: higher fluences of optical excitation create
a thicker layer of the metallized phase. The maximum reflectivity
increase diminishes with temperature. We studied the transient
characteristics of the effect by monitoring the reflected signal
with a photo-detector and a real time digital scope (see Fig.
\ref{Fig3}). The overall bandwidth of the registration system was
125MHz. The transient "switch-on" time has not been resolved in
this experiment. It might have been as short as 4ps, which was the
intrinsic transition time for a transformation from the
$\alpha$-phase to the metallic phase \cite{Rode}. For a given
excitation level, there is a steep increase in the relaxation time
as the temperature of the structure approaches gallium's melting
temperature $T_0 = 29.6^o$C, while relaxation times as short as 20
ns are observed at temperatures below 14$^\circ$C (Fig.
\ref{Fig4}). This can be explained by the fact that the
recrystallization velocity $v$ in turn depends on temperature:
$v\propto(T-T_0)$ \cite{Peteves}, so the closer the system is to
the melting temperature, the longer the time required for the
meta-stable metallized layer to re-crystalize back to the
$\alpha$-phase.

In conclusion, our experiment demonstrates that light-induced
metallization of $\alpha$-Ga under conditions for "resonant" SPP
coupling can lead to a very effective modulation of the reflected
light intensity. This observation also provides a strong
indication that the active plasmonics concept, proposed in ref
\cite{Krasavin}, will indeed provide an efficient technique for
all-optical modulation of SPP signals with a bandwidth of tens of
MHz. It follows from the previous studies of optically induced
metallization of gallium \cite{Zheludev} that the modulation
mechanism is inherently optically broadband and could provide high
modulation contrast for signals in the visible and near infrared
spectral ranges.

The Support of the Engineering and Physical Sciences Research
Council (UK) is gratefully acknowledged.

\end{document}